# SAGES南山1米大视场望远镜g、r、i波段测光数据流量定标

肖凯[1,2], 苑海波[1,2*], 黄博闻[1,2], 徐帅[1,2], 郑捷[3], 李春[3], 范舟[3], 王炜[3], 赵刚[3], 冯国杰[4], 张轩[4], 刘进忠[4], 张若羿[1,2], 杨琳[5], 张余[4], 白春海[4], 牛虎彪[4], 艾力·伊沙木丁[4], 马路[4]

1. 北京师范大学天文与天体物理前沿科学研究所, 北京 102206;
2. 北京师范大学天文系, 北京 100875;
3. 中国科学院国家天文台, 中国科学院光学天文重点实验室, 北京 100101;
4. 中国科学院新疆天文台, 新疆 830011;
5. 北京师范大学人工智能学院, 北京 100875
* 联系人, E-mail: yuanhb@bnu.edu.cn

2023-01-15 收稿, 2023-04-19 修回, 2023-04-20 接受, 2023-04-21 网络版发表
国家自然科学基金(12173007, 12222301)、国家重点研发计划(2019YFA0405503)和中国载人航天工程巡天空间望远镜专项(CMS-CSST-2021-A08, CMS-CSST-2021-A09)资助

**摘要** 均匀且精确的流量定标是大视场测光巡天数据处理的难点和核心所在. 本文借助LAMOST(Large Sky Area Multi-Object Fiber Spectroscopic Telescope) DR7光谱数据、修正后的Gaia EDR3测光数据和Gaia测光丰度数据, 使用基于光谱的SCR(Stellar Color Regression)方法和基于测光的SCR方法(SCR′方法)构建了南山1米大视场望远镜g、r、i波段精度约为10~20毫星等的260多万颗标准星, 进而对SAGES(Stellar Abundance and Galactic Evolution Survey)南山1米望远镜拍摄的约4254平方度天区的测光数据进行了相对流量定标. 在此过程中, 我们仔细考虑了定标零点随不同图像(时间)、探测器不同通道及星像在探测器不同位置的变化. 基于修正后的Pan-STARRS(PS1) DR1测光数据, 进行了绝对流量定标, 并给出了南山g、r、i波段定标后的星等与PS1星等之间的相互转换关系. 基于相邻图之间的共同源对定标精度进行了内部一致性检验, 发现3个波段的定标精度均约为2.0毫星等. 同时, 我们还对定标精度做了外部检验, 使用Gaia DR3无缝光谱合成的PS1星等, 发现定标均匀性在1.3°分辨本领下分别为2.4、2.3和0.9毫星等. 此外, 本文还讨论了恒星平场和探测器通道间相对增益随时间的变化.

当代以及下一代大视场测光巡天项目不胜枚举, 如Sloan Digital Sky Survey(SDSS)[1]、Beijing-Arizona-Taipei-Connecticut(BATC) Sky Survey[2]、South Galactic Cap u-band Sky Survey(SCUSS)[3~5]、Xuyi Schmidt Telescope Photometric Survey of the Galactic Anti-center (XSTPS-GAC)[6]、Beijing-Arizona Sky Survey (BASS)[7]、Stellar Abundance and Galactic Evolution Survey(SAGES)[8,9]、Pan-STARRS1 Survey(PS1)[10]、China Space Station Telescope(CSST)[11]、Wide Field Survey Telescope(WFST)[12]、Multi-channel Photometric Survey Telescope(Mephisto)[13]及司天工程[14]等. 这些项目的开展对恒星物理、银河系结构、星系形成与演化、暗物质和暗能量、时域天文等天文学重大领域的发展具有重要作用.

SAGES采用一套对恒星大气参数敏感的独特滤光片系统：$u_s$、$v_s$、g、r、i、$H\alpha_n$、$H\alpha_w$、$\lambda_{DDO51}$，面向北天约1.2万平方度的高银纬天区(银纬高于10°)展开了测光观测[8]. 其中，g、r、i波段的观测由中国科学院新疆天文台南山1米大视场望远镜完成. 南山1米大视场望远镜是地平式结构，位于新疆南山观测站，其平均大气视宁度约2.0″、全年晴夜数约为220夜. 望远镜的视场为1.3°×1.3°，跟踪精度优于1.25″，指向精度优于3″. 探测器采用E2V蓝敏背照式电荷耦合器(charge coupled device, CCD)，包含4096×4136个像元，4通道读出. 鉴于SDSS已完成了北天大部分天区(大于12000平方度)的观测，南山1米望远镜主要面向北天SDSS覆盖天区之外的约4254平方度天区进行巡天观测.

均匀且精确的流量定标，是影响大视场测光巡天成功与否的关键因素之一，也是大视场测光巡天数据处理的难点所在[15]. 目前，大视场测光巡天流量定标方法可分为3类[15]：经典的标准星方法、硬件驱动/观测驱动的定标方法(Ubercalibration[16]、Hypercalibration[17]、Forward Global Calibration Method[18])及软件驱动/物理驱动的定标方法(Stellar Locus Regression[19]、Stellar Locus[20]、Stellar Color Regression[21]). 硬件驱动/观测驱动方法指基于对大视场成像观测的深入理解而衍生出的流量定标方法；软件驱动/物理驱动方法是基于对天体性质，特别是对恒星颜色的深入认识而衍生出的流量定标方法. 关于大视场测光巡天流量定标方法的详细总结，及对这些方法优势、局限性和后续发展情况的综合评述与讨论可参考文献[15].

SCR(Stellar Color Regression)方法由Yuan等人[21]于2015年提出. 他们将其应用于Stripe 82天区的颜色定标，得到了优于之前结果2~3倍的定标精度[21](在g、r、i波段组成的颜色上精度约为2个毫星等). SCR方法的核心思想是基于对恒星内秉性质的理解，借助已有的观测数据来预测恒星的内秉颜色，继而构建海量颜色标准星[15,22]. 表征恒星内秉性质的具体数据形式可以是大规模光谱巡天得到的恒星大气参数(如有效温度、金属丰度及表面重力加速度等)、恒星光谱、恒星多波段颜色、考虑金属丰度影响的颜色-颜色关系. 通过将海量高精度红化改正后颜色标准星的颜色与实际测量的恒星颜色进行比较，可以有效地改正地球大气及仪器效应带来的各种系统误差.

本文借助LAMOST(Large Sky Area Multi-Object Fiber Spectroscopic Telescope) DR7光谱数据、Gaia EDR3测光数据和Gaia测光金属丰度数据，使用基于光谱的SCR方法和基于测光的SCR方法，分别在3个波段上构建流量标准星，进而对SAGES南山1米望远镜g、r、i波段测光数据进行流量定标.

## 1 数据

### 1.1 SAGES南山数据

SAGES利用南山1米大视场望远镜对北天约4254平方度天区展开了测光观测，在g、r、i波段上分别获得了5283、4849、5314幅测光数据. SAGES南山1米望远镜自2016年8月开始，2018年1月结束. 在g、r、i波段，曝光时间为40 s. 3个波段上10$\sigma$极限星等约为20.0~20.5星等. 典型视宁度(seeing)为2″. 采用专用处理程序对数据进行了初步处理，平场改正采用了晨昏天光平场. 地平式望远镜场旋对平场改正的影响在第4节讨论. 找源和测光采用了业界成熟的软件Source-Extractor[23]. 在定标的过程中，我们采用的是mag_auto星等.

### 1.2 Gaia EDR3测光数据

欧洲航天局Gaia空间卫星[24]EDR3数据[25,26]提供了全天约18亿颗恒星G、$G_{BP}$和$G_{RP}$三个波段的测光数据，定标精度的空间均匀性在毫星等量级. 最近，Yang等人[27]使用结合机器学习的SCR方法，借助约10000颗具有精确*UBVRI*星等测量的Landolt标准星，对Gaia EDR3的测光数据定标系统误差进行了高精度测量，发现其随星等变化可高达10毫星等. 本文使用的Gaia EDR3测光数据已对上述系统误差进行了修正.

### 1.3 LAMOST DR7光谱数据

LAMOST望远镜[28~31]位于国家天文台兴隆观测站，配备了4000条光纤，视场约为20平方度. LAMOST DR7数据[28]包含10640255条低分辨率光谱，波长范围覆盖整个光学波段. 通过LASP(LAMOST Stellar Parameter Pipeline[32])得到恒星的基本参数：有效温度$T_{\rm eff}$、表面重力加速度log*g*和金属丰度[Fe/H]等，其典型精度分别为110 K、0.2 dex和0.1 dex[28].

### 1.4 Gaia测光金属丰度数据

最近，Xu等人[33]借助LAMOST DR7光谱数据和修正后的Gaia EDR3测光数据，估计了高银纬天区($|b| > 10°$，其中$b$为银纬)约2700万颗$10 < G < 16$等恒星

(600余万颗巨星和2000余万颗矮星)的测光金属丰度, 典型误差约为0.2 dex. 该数据为南山数据的流量定标提供了很好的契机.

## 2 方法

本文使用两种方法对SAGES南山1米望远镜g、r、i波段测光数据进行流量定标: 基于光谱的SCR方法和基于测光的SCR方法(SCR′). 基于光谱的SCR方法借助恒星光谱数据展开, 适用于有精确光谱观测的源. 该方法中, 以基于恒星光谱获得的大气参数(有效温度、金属丰度、表面重力加速度等)限制恒星的内秉颜色. 红化改正过程中, 恒星消光值来自结合光谱参数和Gaia测光数据通过恒星配对方法[34,35]得到的较为精确无偏$E(G_{BP}-G_{RP})$, 颜色相对$E(G_{BP}-G_{RP})$的红化系数将通过迭代方式得到. 在没有或缺少光谱数据的情况下, SCR′方法以测光丰度代替光谱得到的金属丰度, 恒星内秉颜色预测来自丰度依赖的恒星颜色轨迹[36]. 红化改正过程中, 恒星消光值来自Schlegel等人[37]的2D消光图$E(B-V)_{SFD}$, 红化系数则可以基于Zhang和Yuan[38]的研究结果得到. 两种方法定标流程如图1所示.

### 2.1 基于光谱的SCR方法(SCR方法)

本文中, 基于光谱的SCR方法(简称SCR方法)是指基于恒星光谱获得的大气参数来限制恒星的内秉颜色. 具体实现过程如下.

($a_1$) 南山1米望远镜g、r、i波段测光数据、LAMOST DR7光谱数据和修正后的Gaia EDR3测光数据进行交叉匹配, 交叉半径为1″.

($b_1$) 选择满足如下条件的主序星($\log g > -3.4\times 10^{-4}\times T_{\rm eff}+5.8$)作为SCR方法的定标样本: (1) g、r、i波段星等误差err{$g$, $r$, $i$}<0.02星等; (2) phot_bp_rp_excess_factor< $1.3+0.06\times(G_{BP}-G_{RP})^2$用以剔除Gaia测光不可靠的源; (3) 4500 K < $T_{\rm eff}$ < 6500 K, [Fe/H] > –1 选择相对窄的温度、丰度范围有利于提高内秉颜色拟合的稳健性; (4) LAMOST光谱在g波段的信噪比$\mathrm{SNR}_g$>20. 最终, g、r、i波段的定标样本分别包含676838、546331和629639颗源. 以g波段为例, 其定标样本在赫罗图上的分布如图S1(a)所示.

($c_1$) 在定标样本中, 首先选择曝光时间和空间位置相邻且标准星个数较多的两幅图, 继而选择两幅图中某个通道上的源作为控制样本. 对应于g、r、i波段, 分别选择文件号(FILENUM)为3354和3355、2433和2434、3006和3007的两幅图. 3个波段控制样本源的个数分别为350、303和338个, 其在空间中均匀分布.

($d_1$) 红化改正环节, 鉴于Schlegel等人[37]的2D消光图$E(B-V)$(以下简称$E(B-V)_{SFD}$)存在空间位置相关的系统误差, 本文使用通过恒星配对方法[36,37]计算得到的 $E(G_{BP}-G_{RP})$. 对应g、r、i波段, 基于Gaia $G_{BP}/G_{RP}$测光数据构建3个颜色: $\boldsymbol{C}=\boldsymbol{G}_{BP,RP}-\boldsymbol{m}^{obs}$. 其中, $\boldsymbol{G}_{BP,RP}=(G_{BP},G_{RP},G_{RP})^{\rm T}$, $\boldsymbol{m}^{obs}=(g,r,i)^{\rm T}$. 接下来, 利用关于温度$T_{\rm eff}$和丰度[Fe/H]的二元二阶多项式(包含6个自由参数, 见式(1))拟合内秉颜色$\boldsymbol{C}_0$:

$$\begin{aligned}\boldsymbol{C}_0^{mod}&=\boldsymbol{a}_0\cdot T_{\rm eff}^2+\boldsymbol{a}_1\cdot[\mathrm{Fe/H}]^2+\boldsymbol{a}_2\cdot T_{\rm eff}\cdot[\mathrm{Fe/H}]\\&+\boldsymbol{a}_3\cdot T_{\rm eff}+\boldsymbol{a}_4\cdot[\mathrm{Fe/H}]+\boldsymbol{a}_5,\end{aligned}\tag{1}$$

本文内秉颜色基于式(2)得到, 其中$\boldsymbol{R}$为3个颜色相对$E(G_{BP}-G_{RP})$的红化系数:

$$\boldsymbol{C}_0=\boldsymbol{C}-\boldsymbol{R}\times E(G_{BP}-G_{RP}),\tag{2}$$

其中, $\boldsymbol{C}=\begin{pmatrix}G_{BP}-g\\G_{RP}-r\\G_{RP}-i\end{pmatrix}$, $\boldsymbol{R}=\begin{pmatrix}R_{G_{BP}-g}\\R_{G_{RP}-r}\\R_{G_{RP}-i}\end{pmatrix}$.

($e_1$) 在利用包含温度$T_{\rm eff}$和丰度[Fe/H]的二元二阶多项式拟合控制样本内秉颜色的过程中发现, 3个波段的拟合残差皆呈现随CCD位置$(X,Y)$高达0.05星等的变化. 接下来, 我们使用关于$(X,Y)$的二元二阶多项式拟合残差, 得到每个波段上控制样本的平场修正项$f(X,Y)$. 平场修正后的结果带回$c_1$步, 执行迭代. 最终得到平场改正后内秉颜色关于温度$T_{\rm eff}$和丰度[Fe/H]的关系式.

($f_1$) 每个波段, 将上一步得到的内秉颜色关系用于定标星, 通过式(3)和(4)得到定标样本观测星等与模型星等之差$\Delta\boldsymbol{m}$:

$$\begin{aligned}\boldsymbol{m}_{SCR}^{mod}&=\boldsymbol{G}_{BP,RP}-\boldsymbol{C}_{0,SCR}^{mod}(T_{\rm eff},[\mathrm{Fe/H}])\\&-\boldsymbol{R}\times E(G_{BP}-G_{RP}),\end{aligned}\tag{3}$$

$$\Delta\boldsymbol{m}=\boldsymbol{m}^{obs}-\boldsymbol{m}_{SCR}^{mod}.\tag{4}$$

之后, 用包含9个参数的二阶多项式拟合每幅图上

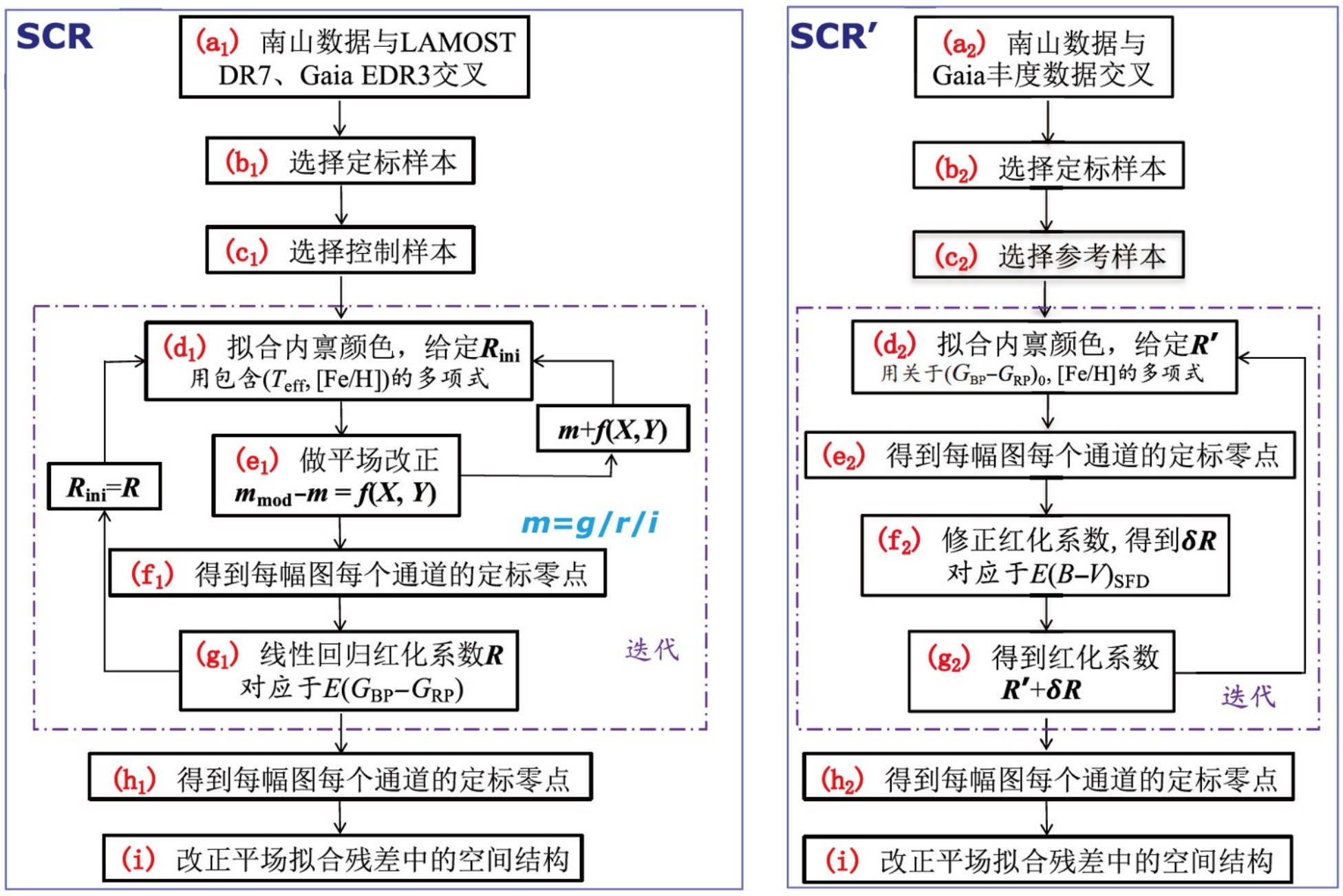

**图 1** 基于光谱的SCR方法和基于测光的SCR方法(SCR′)流程图
**Figure 1** Flowchart of the spectroscopy-based SCR method and the photometric metallicity based SCR method (SCR′) used in this work

的标准星星等差, 即定标零点$ZP_{SCR}$. 二阶多项式形式如下: $\Delta m = a_0X^2 + a_1Y^2 + a_2XY + a_3X + a_4Y + a_5 + a_6\delta(\mathrm{gate}_1) + a_7\delta(\mathrm{gate}_2) + a_8\delta(\mathrm{gate}_3)$. 其中, $\delta(\mathrm{gate}_i)$的值对于第$i$个通道是1, 对于其他通道是0; $X$和$Y$代表源在CCD上的位置. 3个常数项$a_6$、$a_7$和$a_8$分别代表3个通道相对另外一个通道的差值, 该差值反映的是增益的相对变化.

($g_1$) 在每个波段, 零点改正后定标样本的观测颜色与内秉颜色之差即为红化值, $\boldsymbol{R} \times E(G_{BP} - G_{RP}) = \boldsymbol{C} - \boldsymbol{C}_0$. 之后, 通过线性拟合的方式回归红化系数, 并进行了3$\sigma$剔除. 将红化系数带入内秉颜色拟合过程, 执行迭代. 由于$G_{BP}$和$G_{RP}$波段的带通较宽, $G_{BP} - g$和$G_{RP} - r$颜色的红化系数拟合过程考虑了温度依赖性. 线性回归红化系数过程, 强制拟合直线过(0, 0)点. 详细描述见第3节.

($h_1$) 借助迭代后的红化系数, 得到最终定标零点.

(i) 改正平场拟合残差在CCD位置上呈现出的空间结构. 详细描述见第4节.

在SCR方法中, 对每个通道上标准星个数都大于20的图像, 通过多项式拟合的方式得到该图的定标零点. 为了构建更多的标准星以定标更多图, 接下来, 基于测光丰度数据利用基于测光的SCR方法进行流量定标.

## 2.2 基于测光的SCR方法(SCR′方法)

本文中, 基于测光的SCR方法指用丰度依赖的恒星颜色轨迹[36]来限制恒星的内秉颜色. 例如: 用内秉颜色$(G_{BP} - G_{RP})_0$和丰度[Fe/H]限定恒星的内秉颜色. 为了与基于光谱的SCR方法作区分, 基于测光的SCR方法在下文中简称SCR′方法, 流程如下.

($a_2$) 南山g、r、i波段测光数据与Gaia测光丰度数据交叉匹配, 交叉半径为1″.

($b_2$) 选择满足如下条件的矮星作为定标样本: (1) 星等误差err{$g$, $r$, $i$}<0.02星等; (2) phot_bp_rp_excess_factor< $1.3 + 0.06 \times (G_{BP} - G_{RP})^2$; (3) 银盘距大于300 pc, 以保证$E(B-V)_{SFD}$的有效性. 最终, g、r、i波段的定标星个数分别为2632810、2128017和2451425个. g波段的定标样本如图S1(b)所示.

($c_2$) 选择SCR′方法的定标星与SCR方法的共同源作为参考样本.

($d_2$) 与SCR方法类似, 对应于南山数据的g、r、i波段, 借助Gaia $G_{BP}$/$G_{RP}$测光数据构建3个颜色$\boldsymbol{C}'$. 内秉颜色一方面可以通过观测颜色与红化值做差(式(5))

得到:

$$\boldsymbol{C}'_0 = \boldsymbol{C}' - \boldsymbol{R}' \times E(B-V)_{\rm SFD}, \tag{5}$$

另一方面可以通过关于$(G_{\rm BP}-G_{\rm RP})_0$和[Fe/H]的二元多项式得到. 在红化改正环节, 使用的是2D消光图$E(B-V)_{\rm SFD}$. 最近, Zhang和Yuan[38]基于LAMOST DR7光谱数据和多波段测光数据, 使用恒星配对方法精确计算出21个颜色对应于$E(B-V)_{\rm SFD}$的红化系数, 此过程考虑了消光和温度依赖性. 我们通过Zhang和Yuan[38]的研究结果组合出$G_{\rm BP}-g_{\rm PS1}$、$G_{\rm RP}-r_{\rm PS1}$和$G_{\rm RP}-i_{\rm PS1}$对应于$E(B-V)_{\rm SFD}$的红化系数$\boldsymbol{R}'$. 定标星的温度来自肖凯等人的研究结果(未发表).

消光改正后, 利用关于$(G_{\rm BP}-G_{\rm RP})_0$和[Fe/H]的二元二阶多项式拟合内秉颜色$\boldsymbol{C}'_0$. 研究发现, g、r波段的内秉颜色残差随$E(B-V)_{\rm SFD}$呈现近0.01星等的变化, 其原因主要是南山测光系统与Pan-STARRS测光系统的差异. 变化趋势在i波段很小, 可以忽略.

($\mathrm{e_2}$) 将关于$(G_{\rm BP}-G_{\rm RP})_0$和[Fe/H]的多项式带入式(6), 得到SCR′方法的模型星等:

$$\begin{aligned}\boldsymbol{m}^{\rm mod}_{\rm SCR'} = \boldsymbol{G}_{\rm BP,\,RP} - \boldsymbol{C}^{\rm mod}_{0,\rm SCR'}\left((G_{\rm BP}-G_{\rm RP})_0, [\mathrm{Fe/H}]\right) \\ - \boldsymbol{R}' \times E(B-V)_{\rm SFD}.\end{aligned} \tag{6}$$

之后, 用与SCR方法中同样含有9个系数的二阶多项式拟合每幅图上的星等差$\Delta \boldsymbol{m}'$, 进而得到每幅图上对应的多项式系数, 即定标零点$ZR_{\rm SCR'}$.

($\mathrm{f_2}$) 零点改正后, 修正内秉颜色拟合残差随消光的变化: 首先将内秉颜色拟合残差与$E(B-V)_{\rm SFD}$做除, 然后用关于$E(B-V)_{\rm SFD}$的一元四阶多项式拟合比值, 将其作为红化系数修正项$\delta\boldsymbol{R}'$. 最终用于红化系数修正的多项式系数, 如表S1所示.

($\mathrm{g_2}$) 用修正后的红化系数$(\boldsymbol{R}'+\delta\boldsymbol{R}')$重新进行消光改正和内秉颜色拟合, 执行迭代.

($\mathrm{h_2}$) 借助迭代后的红化系数, 得到最终定标零点.

(i) 改正平场拟合残差在CCD位置上呈现出的空间结构. 详细描述见第4节.

值得说明的是, 对于每个通道上标准星个数都大于20的图像, 我们通过多项式拟合的方式得到该图的定标零点(与SCR方法类似); 对于少于20颗源的图(7%), 用图中标准星的模型星等与南山星等之差的中值作为该图的定标零点. 图S2展示了在每幅图每个通道上, SCR′方法定标星数量的分布.

## 3 结果

在SCR方法中, 平场改正前后的控制样本内秉颜色拟合结果分别展示在图2(a), (b)中. 可以看到, 平场改正前, 每个波段上的拟合残差皆明显地呈现出与CCD位置$(X, Y)$相关的高达0.05星等的变化. 平场改正后, 拟合残差与温度、丰度、$X$和$Y$皆无关, 且拟合残差的标准偏差在g、r、i波段分别减小了约0.016、0.019和0.028星等. 3个波段上控制样本平场改正结果如图S3所示. 最终的拟合残差在g、r、i波段上分别为0.011、0.012和0.019星等, 这意味着通过LAMOST光谱数据和Gaia测光数据预言的南山g、r、i星等的精度约为0.01~0.02星等. 最终, SCR方法中控制样本的内秉颜色拟合参数如表S2所示.

3个颜色相对$E(G_{\rm BP}-G_{\rm RP})$红化系数的拟合结果如图3所示. 可以看到, $G_{\rm BP}-g$和$G_{\rm RP}-r$两个颜色的红化系数呈现出明显的温度依赖性. 鉴于此, 我们把SCR方法构建的定标星按温度分成16个区间, 每个区间的温度跨度为500 K, 进而通过线性回归的方式分别得到16个温度区间中的红化系数. 接下来, 我们使用一个包含4个自由参数的三阶多项式, 拟合红化系数随温度的变化关系, 最终的拟合参数如表S3所示. 对于$G_{\rm RP}-i$颜色, 其红化系数随温度的依赖性很弱, 可以忽略.

SCR′方法中控制样本的内秉颜色拟合结果如图4所示. 拟合残差随$(G_{\rm BP}-G_{\rm RP})_0$、[Fe/H]、$E(B-V)_{\rm SFD}$的分布都没有呈现出相关性. 拟合残差的标准差在g、r、i波段分别为0.012、0.012和0.016星等, 这与通过LAMOST光谱数据和Gaia测光数据预测的南山g、r、i标准星精度几乎一致. SCR′方法中控制样本内秉颜色拟合过程的拟合参数如表S4所示.

Sun等人[39]的研究表明, $E(B-V)_{\rm SFD}$存在空间分布不均匀的系统误差(0.01~0.02星等). Xiao和Yuan[40]的研究表明, $E(B-V)_{\rm SFD}$中0.02星等的系统差会使得与南山g、r、i波段类似的Pan-STARRS系统g、r、i模型星等分别变化约1.0、2.0和0.4毫星等. 鉴于此, 我们把通过SCR方法和SCR′方法共同得到的图像定标零点, 按照2:1权重组合在一起, 最终得到每个波段每幅图的定标零点$ZP$, 即$ZP = \frac{2}{3} \times ZP_{\rm SCR} + \frac{1}{3} \times ZP_{\rm SCR'}$. 对于少量只有SCR方法或SCR′方法得到零点的图像, 分别以SCR方

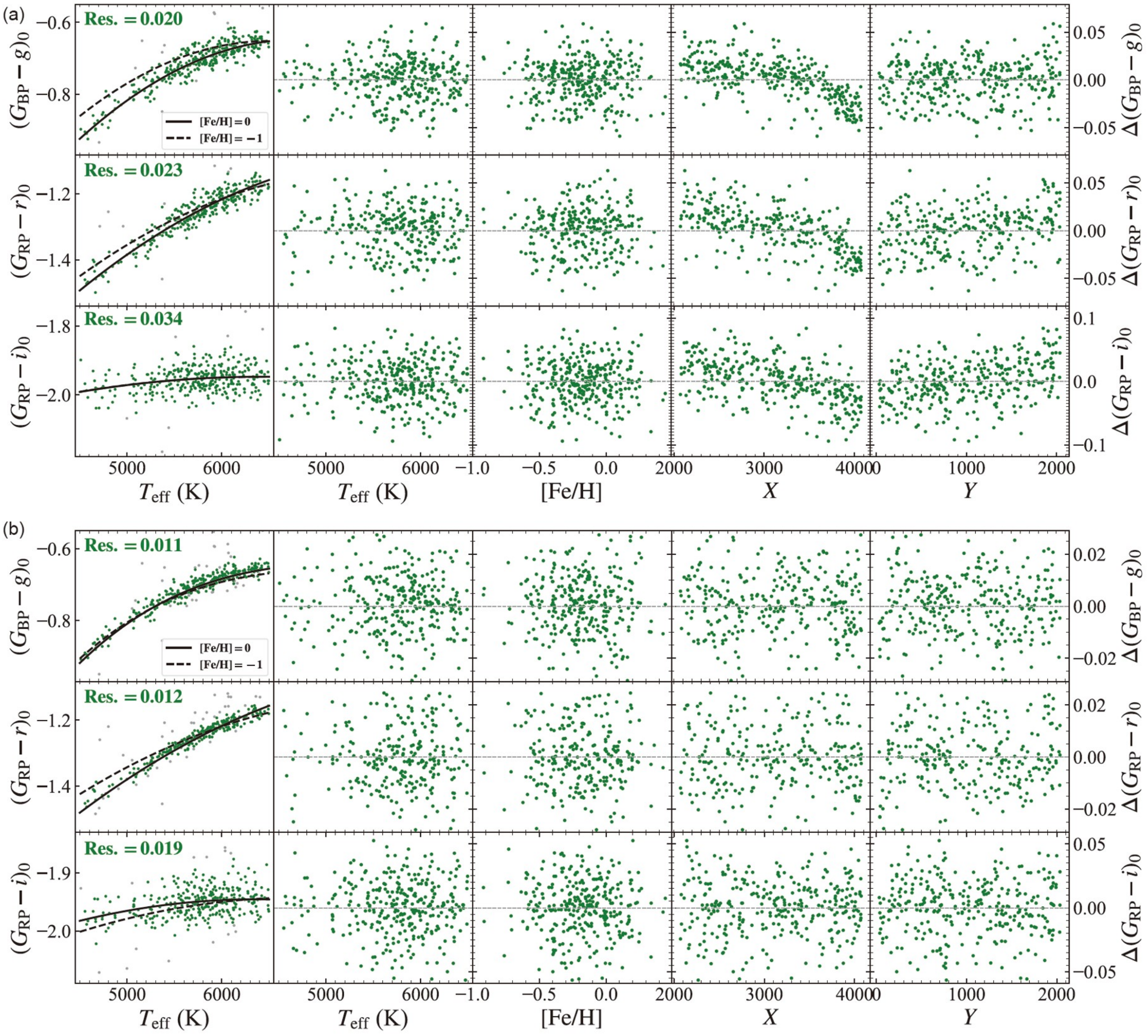

**图 2** 用关于温度和丰度的二元二阶多项式拟合SCR的控制样本内秉颜色，自上而下依次是$G_{BP}-g$、$G_{RP}-r$和$G_{RP}-i$颜色. 自左而右，第一列展示了3$\sigma$剔除后的拟合结果；离群点用灰色加号表示；实线和虚线分别代表[Fe/H] = 0和−1时的结果；Res.为拟合残差. 第二、三、四和五列分别展示了拟合残差随$T_{eff}$、[Fe/H]、$X$和$Y$的分布. (a) 平场改正前的结果. (b) 平场改正后的结果

**Figure 2** Two-dimensional second-order polynomial fitting of intrinsic colors as functions of temperature and metallicity for the SCR control sample stars. From top to bottom are for the $G_{BP}-g$, $G_{RP}-r$ and $G_{RP}-i$ colors, respectively. From left to right, the first column shows the fitting results after 3$\sigma$ clipping; the outliers are indicated by gray pluses; the solid and dashed curves represent results for [Fe/H] = 0 and −1, respectively; Res.: The fitting residuals. The 2nd, 3rd, 4th and 5th columns plot residuals against $T_{eff}$, [Fe/H], $X$ and $Y$, respectively. (a) The results before flat-fielding correction. (b) The results after flat-fielding correction

法或SCR′方法零点作为图像零点.

为了研究定标零点相对于参考图的空间变化，以每幅图上二阶多项式在$(X, Y) = (2000, 2000)$的值作为该图的定标零点常数，其空间分布如图5所示. 定标零点常数的空间变化由大气消光的变化主导，大气消光大的天区巡天深度浅，因此图5也反映了巡天深度的空间变化.

最近，Xiao和Yuan[40]借助LAMOST数据和Gaia数据，使用3种方法对Pan-STARRS DR1(下文简称PS1)测光数据中存在的随星等和空间变化的系统差做了修正. 本研究中，我们基于修正后的PS1测光数据，对南山1米望远镜g、r、i波段测光数据进行了绝对定标. 在该过程中，我们通过颜色-颜色图(如$g-g_{PS1}$颜色-$g_{PS1}-i_{PS1}$颜色)获得了南山g、r、i星等相对于PS1系统

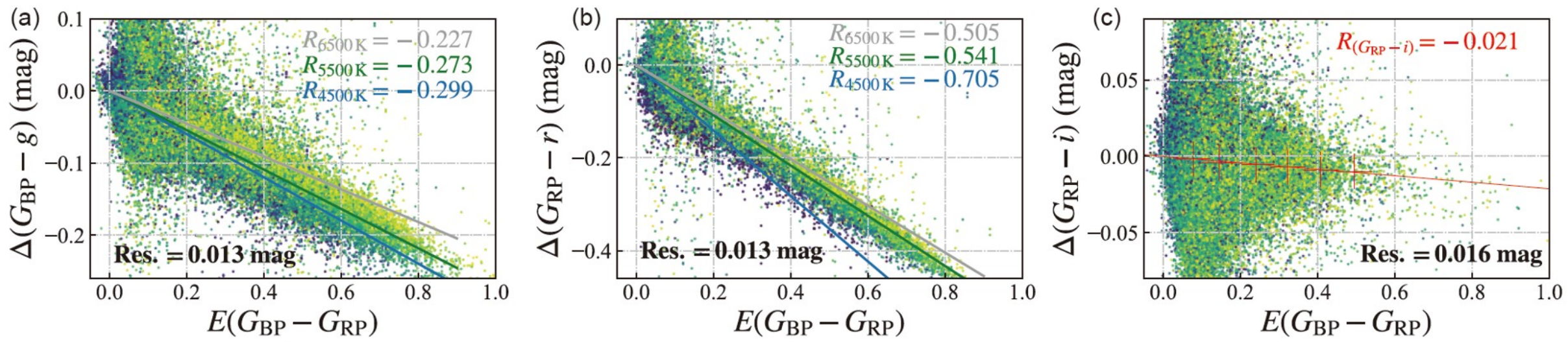

**图 3** 针对SCR定标样本，线性回归$G_{\rm BP}-g$、$G_{\rm RP}-r$和$G_{\rm RP}-i$三个颜色相对$E(G_{\rm BP}-G_{\rm RP})$的红化系数. 为了避免数据点拥挤, 均匀选取了2%的点, 不同点的颜色代表不同的温度. (a), (b) 红化系数随温度有一定程度的相关性, 给出了3个温度区间的红化系数. (c) 红化系数对温度的依赖非常弱. 红色加号为3$\sigma$剔除后的中值; 红色实线是对红色加号拟合的结果

**Figure 3** The linear regression of reddening coefficients of the $G_{\rm BP}-g$, $G_{\rm RP}-r$ and $G_{\rm RP}-i$ colors with respect to $E(G_{\rm BP}-G_{\rm RP})$ for the SCR calibration samples. To avoid crowding, only one in fifty stars are plotted. Colors indicate different temperatures. (a), (b) Reddening coefficients show moderate temperature dependence; the fitted reddening coefficients for three temperature ranges are shown. (c) Reddening coefficients show weak dependence on temperature. The red pluses are the median values after 3$\sigma$ clipping; the red solid lines are linear fits to the red pluses

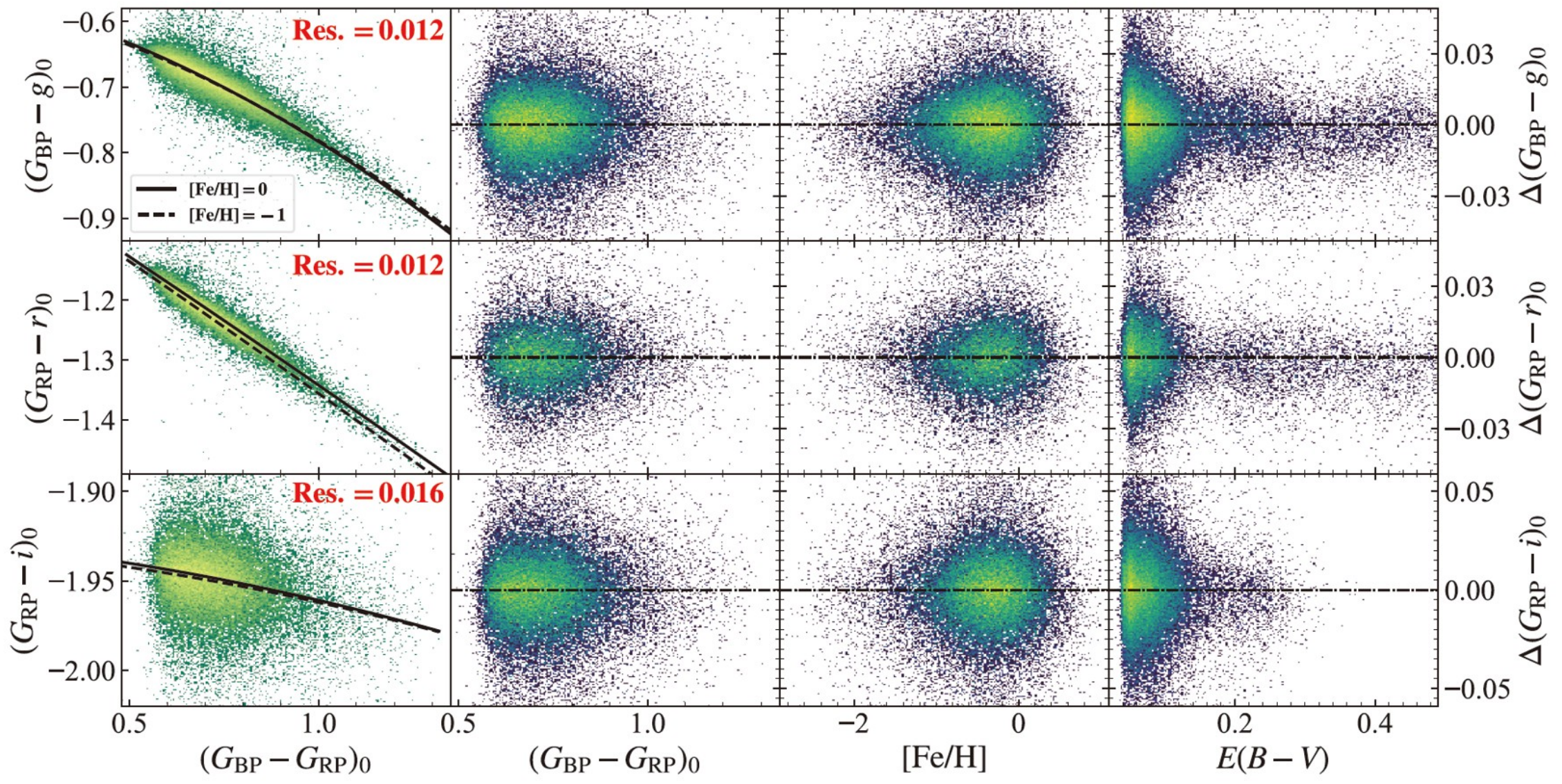

**图 4** 针对SCR′控制样本星，用包含颜色$(G_{\rm BP}-G_{\rm RP})_0$和丰度[Fe/H]的二元二阶多项式拟合控制样本的内秉颜色，自上而下依次是$G_{\rm BP}-g$、$G_{\rm RP}-r$和$G_{\rm RP}-i$颜色. 自左而右, 第一列展示了3$\sigma$剔除后的拟合结果; 离群点用灰色加号表示; 实线和虚线分别代表[Fe/H] = 0和−1时的结果. 第二、三、四列分别展示了拟合残差随$(G_{\rm BP}-G_{\rm RP})_0$、[Fe/H]和$E(B-V)_{\rm SFD}$的变化

**Figure 4** Two-dimensional second-order polynomial fitting of intrinsic colors as functions of $(G_{\rm BP}-G_{\rm RP})_0$ and [Fe/H] for the SCR′ control sample stars. From top to bottom are for the $G_{\rm BP}-g$, $G_{\rm RP}-r$ and $G_{\rm RP}-i$ colors, respectively. From left to right, the first column shows the fitting results after 3$\sigma$ clipping; the outliers are indicated by gray pluses; the solid and dashed curves represent results for [Fe/H] = 0 and −1, respectively. The 2nd, 3rd, and 4th columns plot residuals against $(G_{\rm BP}-G_{\rm RP})_0$, [Fe/H] and $E(B-V)_{\rm SFD}$, respectively

的绝对零点. 该零点在每个波段为常数(−0.660, −0.788, −1.603).

最后, 我们借助低消光样本的颜色-颜色关系, 给出了南山g、r、i波段定标后的星等$g_{\rm NS}$、$r_{\rm NS}$和$i_{\rm NS}$与修正后PS1的$g_{\rm PS1}$、$r_{\rm PS1}$和$i_{\rm PS1}$星等之间的转换关系. 从南山(NS)星等到PS1星等的转换关系为

$$
\begin{aligned}
g_{\rm PS1} &= g_{\rm NS}-0.08\times(g_{\rm NS}-i_{\rm NS})+0.040,\\
r_{\rm PS1} &= r_{\rm NS}-0.03\times(g_{\rm NS}-i_{\rm NS})+0.015,\\
i_{\rm PS1} &= i_{\rm NS}-0.01\times(g_{\rm NS}-i_{\rm NS})+0.010.
\end{aligned} \tag{7}
$$

从PS1星等到NS星等的转换关系为

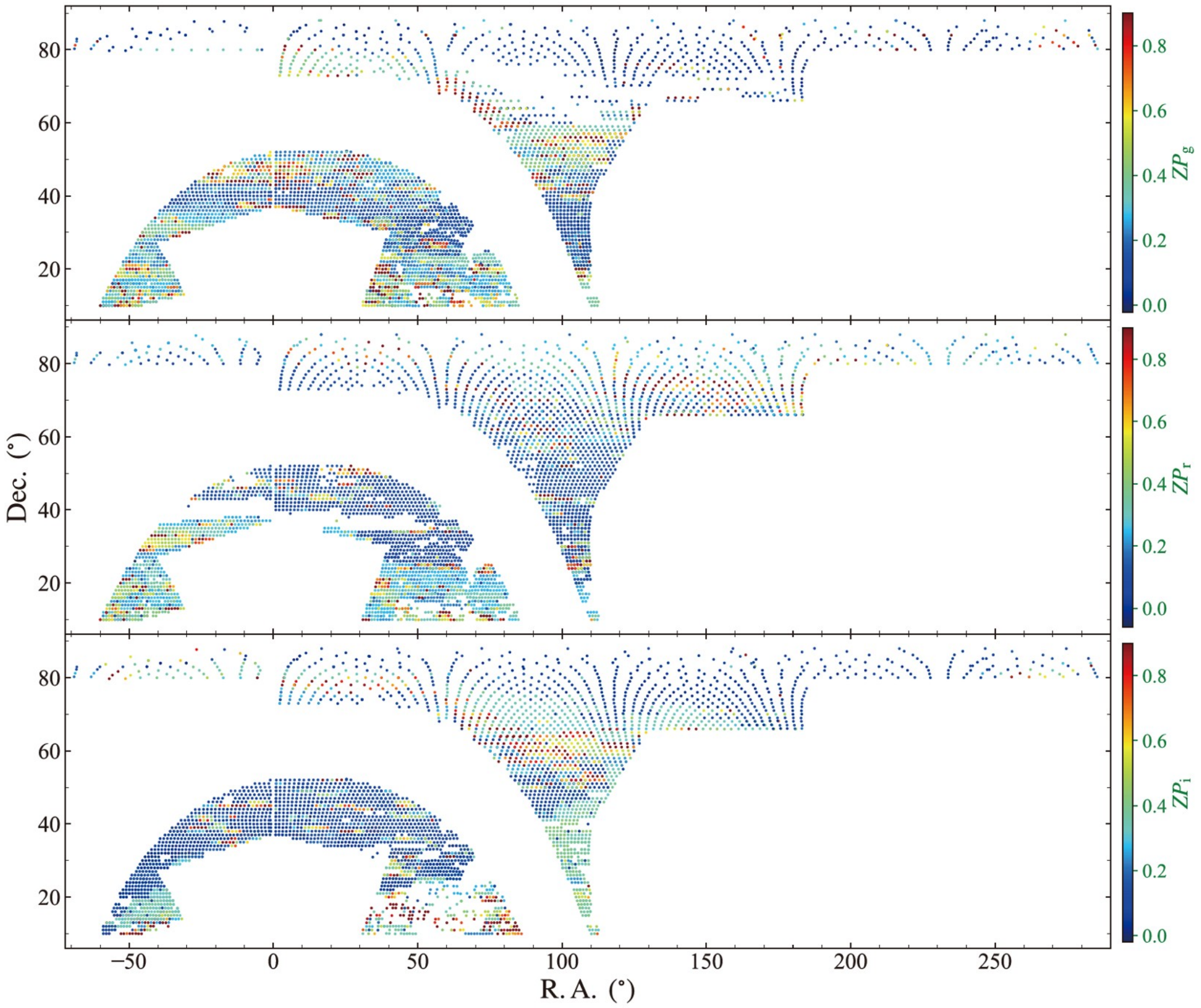

**图 5** 每幅图相对于参考图定标零点常数的空间变化. R.A.表示赤经; Dec.表示赤纬. 自上而下依次是g、r和i波段

**Figure 5** Spatial variations of the calibration zero-points of each image relative to the reference image. R.A. denotes right ascension; Dec. denotes declination. From top to bottom are for the g, r, and i bands, respectively

$$
\begin{aligned}
g_{\mathrm{NS}} &= g_{\mathrm{PS1}} + 0.08 \times (g_{\mathrm{PS1}} - i_{\mathrm{PS1}}) - 0.040, \\
r_{\mathrm{NS}} &= r_{\mathrm{PS1}} + 0.03 \times (g_{\mathrm{PS1}} - i_{\mathrm{PS1}}) - 0.015, \\
i_{\mathrm{NS}} &= i_{\mathrm{PS1}} + 0.01 \times (g_{\mathrm{PS1}} - i_{\mathrm{PS1}}) - 0.010.
\end{aligned} \tag{8}
$$

# 4 讨论

## 4.1 两种方法结果比较

基于两种方法构建的标准星的模型星等与定标后的南山g、r、i星等之差分别随消光、定标后的星等和颜色的变化, 如图S4(a), (b)所示. 与预期相符, 基于SCR方法构建的流量标准星的模型星等与定标后的南山g、r、i星等之差不随消光、定标后的星等和颜色变化而变化; 基于SCR′方法构建的流量标准星的模型星等与定标后的星等之差与消光和颜色无关, 展现出的对定标后星等的微弱依赖(毫星等量级)可以忽略.

我们对两种方法得到的结果进行比较. 基于两种方法所构建的标准星共同源, 图6(a)展示了3个波段上两种方法定标后的星等之差分别随定标后星等、颜色及$E(B-V)_{\mathrm{SFD}}$系统误差$\Delta E(B-V)_{\mathrm{SFD}}$的分布. 其中, $\Delta E(B-V)_{\mathrm{SFD}}$来自Sun等人[39]的研究结果. 与预期相符, 结果与星等和颜色皆无关. 我们注意到, g和r波段的星等差与$\Delta E(B-V)_{\mathrm{SFD}}$皆存在相关性, 这是由SCR′方法红化改正过程中$E(B-V)_{\mathrm{SFD}}$空间相关的系统误差导致的.

为了定量描述两种方法结果的一致性, 我们得到了3个波段上每幅图的零点差随图中SCR′方法源个数的分布, 如图6(b)所示. 通过高斯拟合得到不同源个数区间内零点差的标准偏差. 随着图中源个数的增多, 标准偏差先减小, 后逐渐变平, 最终趋于收敛. g、r、i波段的收敛值约为2.3、2.2和1.9毫星等, 这意味着两种方法的一致性约为2毫星等.

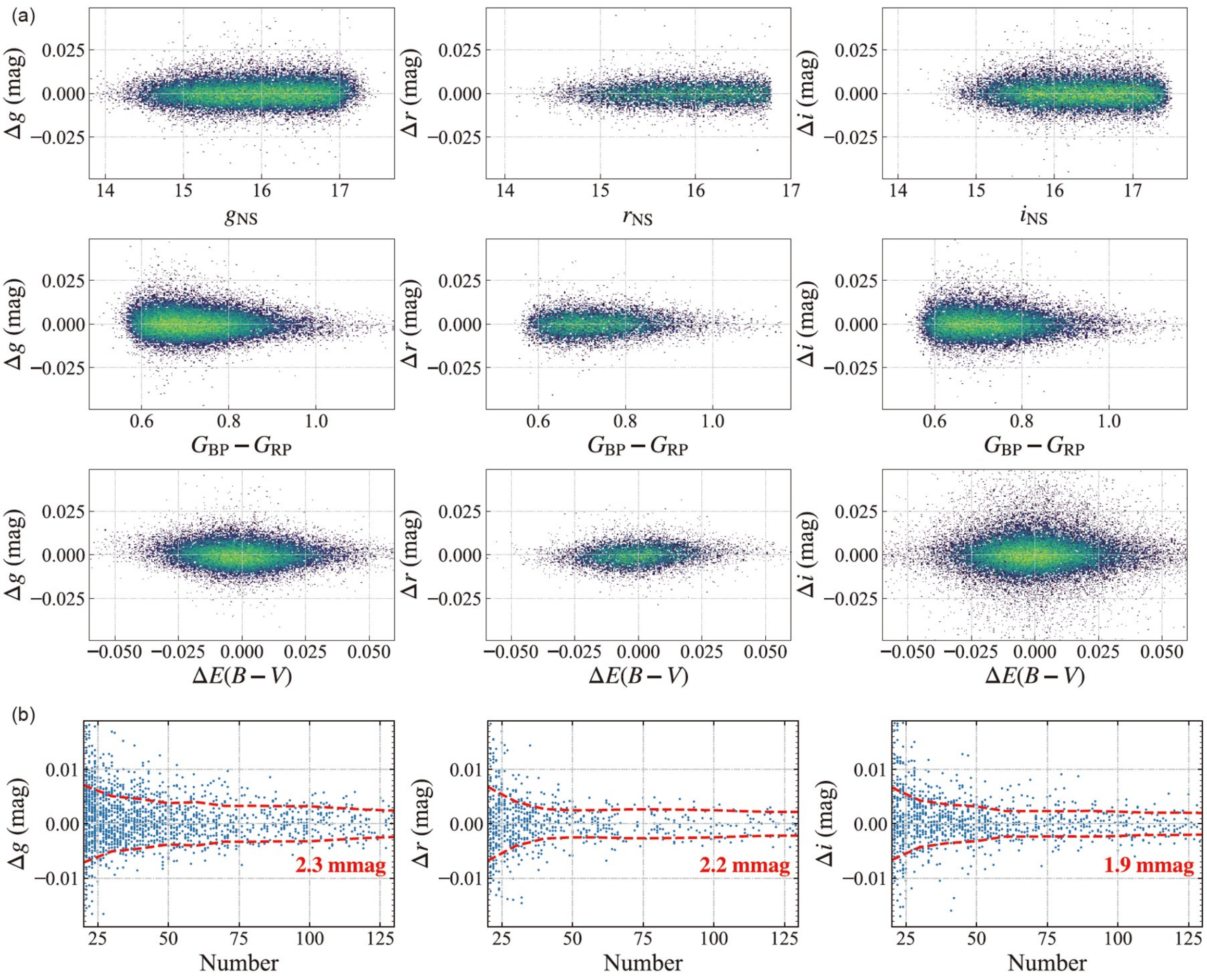

**图 6** SCR和SCR′方法结果的比较. 自左而右依次是g、r、i波段. (a) SCR和SCR′方法得到的星等之差随定标后星等、颜色和$E(B-V)_{SFD}$系统误差的变化. (b) 每幅图上两种方法零点之差随图中两种方法定标星共同源个数的变化. 虚线为标准偏差. 收敛值在图中标出

**Figure 6** Comparison of the results between the SCR and SCR′ methods. From left to right are for the g, r, and i bands, respectively. (a) Differences of derived magnitudes between the SCR and SCR′ methods as a function of calibrated magnitude, color and systematic errors of $E(B-V)_{SFD}$. (b) Differences of zero points between the SCR and SCR′ methods as a function of numbers of common calibration stars between the two methods in one image. The standard deviations are indicated by dashed lines. The convergence value is marked in each panel

## 4.2 恒星平场随时间和空间位置的变化

我们研究了恒星平场随时间和空间位置的变化. 以2017年10月12日连续拍摄的g波段47幅图为例, 图7(a)展示了在47幅图上SCR′方法构建的标准星的模型星等与南山g波段观测星等之差随CCD位置的空间变化. 此处, 我们调平了不同通道间的增益差. 每幅图上标准星个数大于500颗.

二阶多项式重构出的恒星平场如图7(b)所示, 可以看出恒星平场随时间变化而变化. 这种变化可以分为两类: 一类与望远镜指向在天空中的突然变化有关, 例如序号8、9之间, 14、15之间和17、18之间, 这些变化可能与望远镜性能有关(如重力弯沉等). 另一类与望远镜指向无关, 如序号为10的平场和序号为9、11的平场之间的变化. 相比二阶多项式重构出的恒星平场(图7(b)), 离散的恒星平场(图7(a))能更为直接地反映恒星平场的变化. 这是因为图7(a)明显存在传统的低阶多项式拟合无法完美重构的较小尺度结构. 在流量定标过程中, 我们分析了这些天光平场和二阶多项式恒星平场无法改正的结构(图7(c)), 并进行了相应的处理(详细描述见第4.3节).

通过对模型星等与观测星等之差进行无参数化线性插值处理, 得到的恒星平场随天区的变化如图7(d)所示. 该图呈现了地平式南山1米望远镜工作时指向与旋

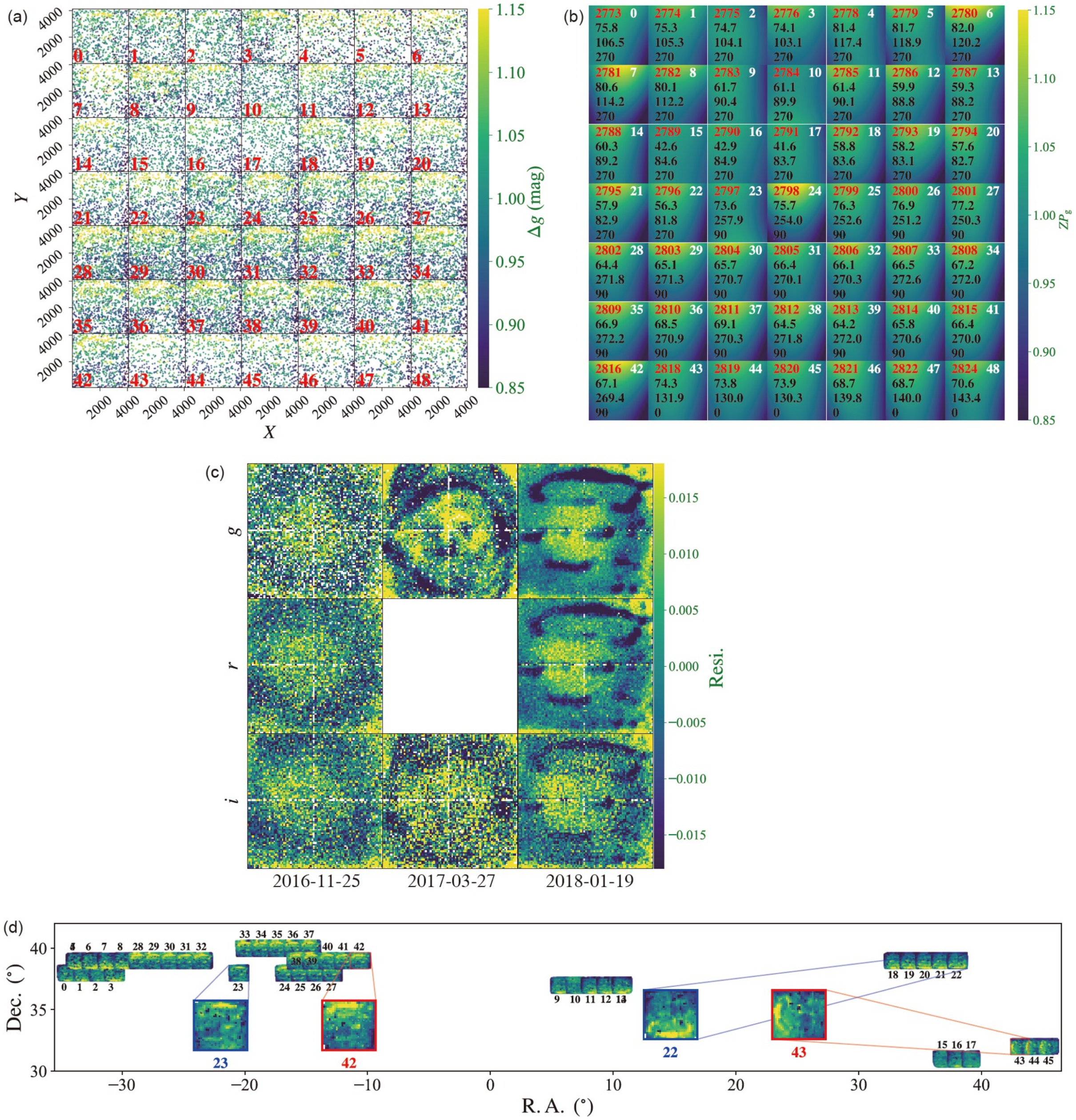

**图 7** 以2017年10月12日g波段曝光的47幅图为例, 展示恒星平场的变化. (a) 预测的星等与NS星等之差在CCD空间上的变化. 子图中序号按观测顺序依次增大, 其与(b)中白色数字及(d)中数字对应. (b) 恒星平场. 每幅图中的数字自上而下分别是FILENUM、序号(白色)、高度角、方位角和旋转角. (c) 举例展示拟合残差在CCD上的空间分布. 从上至下依次是g、r、i波段, 自左而右观测日期分别为2016-11-25、2017-03-27和2018-01-19. (d) 恒星平场的空间分布. 每幅图中的数字为序号, 颜色条与(b)相同

**Figure 7** An example showing the variations of star flats of 47 files observed on October 12, 2017 in the g band. (a) CCD spatial variations of the differences between the derived magnitudes and Nanshan magnitudes. The serial numbers, increasing in the order of observation, correspond to the white numbers in (b) and numbers in (d). (b) Stellar flats. From top to bottom, the FILENUM, serial number (white), elevation angle, telescope azimuth and rotation angle are labeled for each flat. (c) An example showing the spatial pattern of fitting residuals in the CCD plane. From top to bottom are for the g, r, and i bands, and from left to right are for observation dates of 2016-11-25, 2017-03-27 and 2018-01-19. (d) Spatial variations of star flats. For each flat, the serial number is marked, and the color bar matches that of (b)

转的情况, 例如: 序号为22和23的两幅图之间旋转了180°、序号为42和43的两幅图之间旋转了90°等. 同时可以看到, 对于标准星个数不是足够多的图, 无参数化线性插值会放大其恒星平场中的较小尺度结构.

### 4.3 恒星平场拟合残差的空间结构

为了检查零点拟合残差在CCD位置上的空间分布，我们把同一天拍摄的所有图像的拟合残差叠加在一起，发现了3种不同的空间结构，如图S5所示．图S5(a)展示的是不同观测时间下g波段的结果．g波段在18个月共有66天的观测数据．相邻观测时间的残差空间结构几乎是相同的，两次变化都发生在长期(分别约3个月、6个月)巡天观测停止后再次观测之时．其他两个波段结果与g波段类似．

以每个波段在2016-11-25、2017-03-27(当晚没有r波段的观测数据)和2018-01-19三天的观测为例，图7(c)更为清晰地展示了这3种空间结构．可以看到，不同天相同波段之间恒星平场拟合残差的空间结构相似；同一波段在不同时间段，其拟合残差呈现出不同的空间结构．该3种空间结构皆是大尺度平场的特征，其在天光平场改正过程中没有被很好地消除，并且在恒星平场改正过程中无法用低阶多项式进行很好的改正．

为了对平场改正残差在CCD位置上呈现出的空间结构进行改正，我们将每一天拍摄图像上的拟合残差合并在一起，在经过10×10像元合并后，用线性插值的方式对每幅图上经过定标后的源进行改正．

### 4.4 不同通道间相对增益随时间的变化

在SCR和SCR′方法中，我们注意到每幅图上标准星的模型星等与南山g、r、i星等之差在CCD四个通道之间存在明显的系统差．以2016年10月9日拍摄的g波段47幅图为例，图8(a)展示了增益的相对变化．可以看到，模型星等与南山g、r、i星等之差不仅在CCD四个通道之间存在明显的差异，并且该差异随时间变化而变化. 这意味着, 由恒星得到的不同通道间相对增益是随时间变化的．3个通道相对于另一通道的增益变化幅度$\sigma$值约为0.5%~0.7%，在每个通道上最大约为4%．由图8(b)可以看出，经过恒星平场得到的增益进行调平后，每幅图上模型星等与南山g、r、i星等之差在4个通道间连续分布．

### 4.5 定标精度检验

为了定量描述定标的内部一致性，对于每个波段，我们借助相邻两幅图上的重叠源进行如下检验：分别用两幅图上的定标零点确定重叠源定标后的星等，然后用高斯拟合的方式得到任意两幅图上共同源定标后星等差的中值. 接下来, 我们得到了中值随两幅图上共同源个数的变化(图S6), 通过高斯拟合的方式得到不同共同源个数区间内中值的标准偏差. 我们发现, 随着两幅图上共同源个数增大, 标准偏差先快速减小, 后收敛于某个特定值．该收敛值反映了流量定标的内部一致性, 在3个波段上分别为1.7、2.0和2.0毫星等.

本文还对南山测光数据流量定标的一致性开展了外部检验. 首先, 借助基于Gaia DR3 BP/RP(简称XP)无缝光谱[41]的光度合成方法(XP synthetic photometry, XPSP方法[42])合成PS1星等, 继而通过PS1-NS转换关系将PS1星等转换为南山星等$g_{\rm NS}^{\rm PS1}$、$r_{\rm NS}^{\rm PS1}$和$i_{\rm NS}^{\rm PS1}$, 并检验其与定标后的南山星等$g_{\rm NS}$、$r_{\rm NS}$和$i_{\rm NS}$之差的空间分布．对于每个波段，用每幅图上源两者之差的中值作为此天区的星等差, 其空间分布如图S7所示. 通过高斯拟合的方式得到g、r、i波段每幅图上星等差的分布直方图标准偏差，标准偏差分别为2.4、2.3和0.9毫星等．这意味着，与光度合成的PS1星等相比，南山g、r、i波段测光数据的流量定标均匀性在1.3°的天区间约为1~2毫星等．我们定量研究了$E(B-V)_{\rm SFD}$系统误差对SCR′零点的影响．研究发现，$E(B-V)_{\rm SFD}$中0.02星等的系统误差会导致SCR′方法在g和r波段的零点皆变化2个毫星等，但方向相反．故图S7中g、r波段呈现出的空间结构主要由SCR′方法中$E(B-V)_{\rm SFD}$空间相关的系统误差导致．

## 5 总结

本文借助LAMOST DR7光谱数据和修正后的Gaia EDR3测光数据，使用基于光谱的SCR方法构建了60多万颗精度在g、r、i波段分别约为0.011、0.012和0.019星等的流量标准星；借助Gaia丰度数据，使用基于测光的SCR方法(SCR′方法)构建了200多万颗精度在g、r、i波段分别约为0.012、0.012和0.016星等的流量标准星；进而通过使用关于CCD位置坐标$(X, Y)$的二元多项式，分别拟合每幅图上标准星的仪器星等与模型星等之差，得到了每幅图上的定标零点(多项式系数).

本文讨论了恒星平场和4个通道间增益相对值的变化. 研究发现, 恒星平场和通道间相对增益随时间的变化而变化．此外，我们发现，恒星平场零点拟合残差在CCD位置上的空间分布存在3种不同的空间结构．该结构在不同天、相同波段之间是相似的，在同一波段、长时间段跨度(季量级)情况下不尽相同．最终，基于每一天的恒星平场拟合残差的空间结构，用线性插

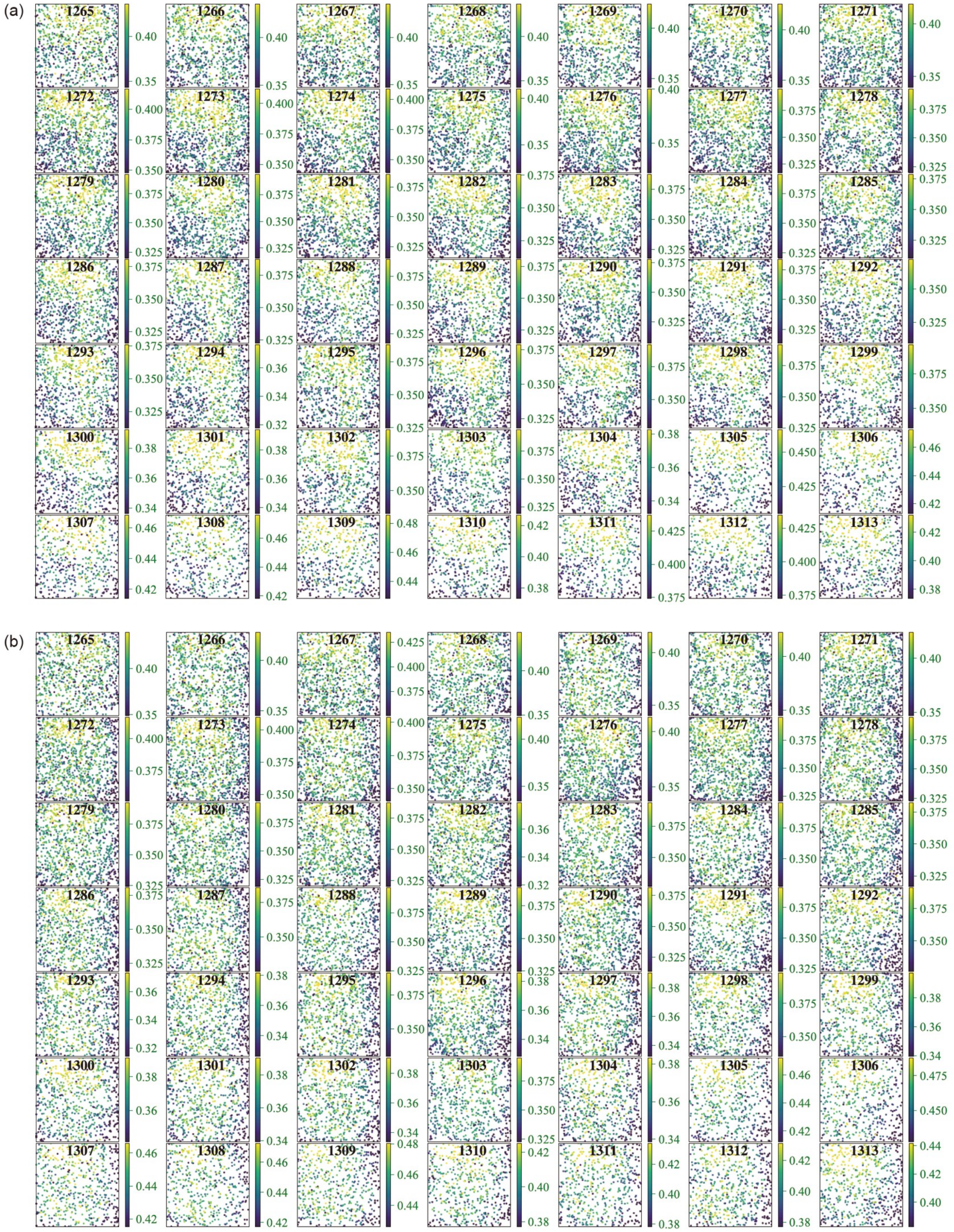

**图 8**　以2016年10月9日g波段曝光的47幅图为例展示不同通道间相对增益的变化. FILENUM(黑色)在每幅子图中标出. (a) SCR′方法预测的星等与NS星等之差. (b) 与(a)类似, 但展示的是增益改正后的结果

**Figure 8**　An example showing the variations of the relative gain values between different amplifiers. A number of 47 files observed on October 9, 2016 in the g band are plotted. For each panel, the FILENUM (black) are marked. (a) Differences between the derived magnitudes by the SCR′ method and the NS magnitudes. (b) Same as (a) but after corrections of gain

值的方式对每一天内所有图做了进一步修正.

基于修正后的Pan-STARRS DR1测光数据, 对南山1米望远镜g、r、i波段测光数据进行了绝对定标, 并给出南山g、r、i波段定标后的星等$g_{NS}$、$r_{NS}$和$i_{NS}$与修正

后PS1的$g_{\mathrm{PS1}}$、$r_{\mathrm{PS1}}$和$i_{\mathrm{PS1}}$星等之间的相互转换关系.

本文基于每幅图与相邻图之间的共同源对流量定标进行了内部一致性检验. 通过比对相邻图之间共同源定标后星等, 我们发现, 流量定标的内部一致性在3个波段分别为1.7、2.0和2.0毫星等. 另外, 基于借助XPSP方法使用Gaia DR3无缝光谱合成的PS1星等, 本文还对南山g、r、i波段测光数据流量定标的一致性做了外部检验. 我们发现, 与光度合成的PS1星等相比, 南山g、r、i波段测光数据的流量定标均匀性在1.3°的分辨本领下分别为2.4、2.3和0.9毫星等.

## 参考文献

1 York D G, Adelman J, Anderson Jr J E, et al. The Sloan Digital Sky Survey: Technical summary. Astron J, 2000, 120: 1579–1587
2 Fan X, Burstein D, Chen J S, et al. Deep wide-field spectrophotometry of the open cluster M67. Astron J, 1996, 112: 628
3 Zhou X, Fan X H, Fan Z, et al. South Galactic Cap u-band Sky Survey (SCUSS): Project overview. Res Astron Astrophys, 2016, 16: 017
4 Zou H, Zhou X, Jiang Z, et al. South Galactic Cap u-band Sky Survey (SCUSS): Data release. Astron J, 2016, 151: 37
5 Zou H, Jiang Z, Zhou X, et al. South Galactic Cap u-band Sky Survey (SCUSS): Data reduction. Astron J, 2015, 150: 104
6 Liu X W, Yuan H B, Huo Z Y, et al. LSS-GAC-A LAMOST spectroscopic survey of the Galactic anti-center. Proc IAU, 2014, 298: 310–321
7 Zou H, Zhou X, Fan X, et al. Project overview of the Beijing-Arizona Sky Survey. Publ Astron Soc Pac, 2017, 129: 064101
8 Zheng J, Zhao G, Wang W, et al. The SAGE photometric survey: Technical description. Res Astron Astrophys, 2018, 18: 147
9 Zheng J, Zhao G, Wang W, et al. Test area of the SAGE survey. Res Astron Astrophys, 2019, 19: 003
10 Schlafly E F, Finkbeiner D P, Jurić M, et al. Photometric calibration of the first 1.5 years of the Pan-STARRS1 survey. Astrophys J, 2012, 756: 158
11 Zhan H. The wide-field multiband imaging and slitless spectroscopy survey to be carried out by the Survey Space Telescope of China Manned Space Program (in Chinese). Chin Sci Bull, 2021, 66: 1290–1298 [詹虎. 载人航天工程巡天空间望远镜大视场多色成像与无缝光谱巡天. 科学通报, 2021, 66: 1290–1298]
12 Lou Z, Liang M, Yao D, et al. Optical design study of the wide field survey telescope (WFST). SPIE, 2016, 10154: 101542A
13 Yuan X, Li Z, Liu X, et al. Development of the multi-channel photometric survey telescope. SPIE, 2020, 11445: 114457M
14 Liu J, Soria R, Wu X F, et al. The SiTian Project. An Acad Bras Ciênc, 2021, 93: 20200628
15 Huang B W, Xiao K, Yuan H B. Photometric calibration methods for wide-field photometric surveys (in Chinese). Sci Sin-Phys Mech Astron, 2022, 52: 289503 [黄博闻, 肖凯, 苑海波. 大视场测光巡天流量定标方法. 中国科学: 物理学 力学 天文学, 2022, 52: 289503]
16 Padmanabhan N, Schlegel D J, Finkbeiner D P, et al. An improved photometric calibration of the sloan digital sky survey imaging data. Astrophys J, 2008, 674: 1217–1233
17 Finkbeiner D P, Schlafly E F, Schlegel D J, et al. Hypercalibration: A Pan-STARRS1-based recalibration of the sloan digital sky survey photometry. Astrophys J, 2016, 822: 66
18 Burke D L, Rykoff E S, Allam S, et al. Forward global photometric calibration of the dark energy survey. Astron J, 2018, 155: 41
19 High F W, Stubbs C W, Rest A, et al. Stellar locus regression: Accurate color calibration and the real-time determination of galaxy cluster photometric redshifts. Astron J, 2009, 138: 110–129
20 López-Sanjuan C, Varela J, Cristóbal-Hornillos D, et al. J-PLUS: Photometric calibration of large-area multi-filter surveys with stellar and white dwarf loci. Astron Astrophys, 2019, 631: A119
21 Yuan H, Liu X, Xiang M, et al. Stellar color regression: A spectroscopy-based method for color calibration to a few millimagnitude accuracy and the recalibration of stripe 82. Astrophys J, 2015, 799: 133
22 Huang B, Yuan H. Photometric recalibration of the SDSS Stripe 82 to a few millimagnitude precision with the stellar color regression method and Gaia EDR3. Astrophys J Suppl Ser, 2022, 259: 26
23 Bertin E, Arnouts S. Extractor: Software for source extraction. Astron Astrophys, 1996, 117(Suppl): 393–404
24 Prusti T, de Bruijne J H J, Brown A G A, et al. The Gaia mission. Astron Astrophys, 2016, 595: A1
25 Riello M, De Angeli F, Evans D W, et al. Gaia Early Data Release 3. Summary of the contents and survey properties. Astron Astrophys, 2021, 649: A3
26 Riello M, De Angeli F, Evans D W, et al. Gaia Early Data Release 3. Summary of the contents and survey properties (Corrigendum). Astron Astrophys, 2021, 650: C3
27 Yang L, Yuan H, Zhang R, et al. Correction to the photometric magnitudes of the Gaia Early Data Release 3. Astrophys J Lett, 2021, 908: L24
28 Luo A L, Zhao Y H, Zhao G, et al. The first data release (DR1) of the LAMOST regular survey. Res Astron Astrophys, 2015, 15: 1095–1124
29 Deng L C, Newberg H J, Liu C, et al. LAMOST Experiment for Galactic Understanding and Exploration (LEGUE)—The survey's science plan. Res Astron Astrophys, 2012, 12: 735–754 30 Zhao G, Zhao Y H, Chu Y Q, et al. LAMOST spectral survey—An overview. Res Astron Astrophys, 2012, 12: 723–734
31 Cui X Q, Zhao Y H, Chu Y Q, et al. The Large Sky Area Multi-Object Fiber Spectroscopic Telescope (LAMOST). Res Astron Astrophys, 2012, 12: 1197–1242
32 Wu Y, Luo A L, Li H N, et al. Automatic determination of stellar atmospheric parameters and construction of stellar spectral templates of the Guoshoujing Telescope (LAMOST). Res Astron Astrophys, 2011, 11: 924–946
33 Xu S, Yuan H, Niu Z, et al. Stellar loci. V. Photometric metallicities of 27 million FGK stars based on Gaia Early Data Release 3. Astrophys J Suppl Ser, 2022, 258: 44
34 Yuan H B, Liu X W, Xiang M S. Empirical extinction coefficients for the GALEX, SDSS, 2MASS and WISE passbands. Mon Not Roy Astron Soc, 2013, 430: 2188–2199
35 Zhang Y, Xu Z, Zhang Q, et al. Sunspot shearing and sudden retraction motion associated with the 2013 August 17 M3.3 flare. Astrophys J Lett, 2022, 933: L20
36 Yuan H, Liu X, Xiang M, et al. Stellar loci. I. Metallicity dependence and intrinsic widths. Astrophys J, 2015, 799: 134
37 Schlegel D J, Finkbeiner D P, Davis M. Maps of dust infrared emission for use in estimation of reddening and cosmic microwave background radiation foregrounds. Astrophys J, 1998, 500: 525–553
38 Zhang R, Yuan H. Empirical temperature- and extinction-dependent extinction coefficients for the GALEX, Pan-STARRS 1, Gaia, SDSS, 2MASS, and WISE passbands. Astrophys J Suppl Ser, 2023, 264: 14
39 Sun Y, Yuan H, Chen B. Validations and corrections of the SFD and Planck reddening maps based on LAMOST and Gaia Data. Astrophys J Suppl Ser, 2022, 260: 17
40 Xiao K, Yuan H. Validation and improvement of the Pan-STARRS photometric calibration with the stellar color regression method. Astrophys J, 2022, 163: 185
41 Carrasco J M, Weiler M, Jordi C, et al. Internal calibration of Gaia BP/RP low-resolution spectra. Astron Astrophys, 2021, 652: A86
42 Gaia Collaboration, Montegriffo P, Bellazzini M, et al. Gaia Data Release 3: The Galaxy in your preferred colours. Synthetic photometry from Gaia low-resolution spectra. 2022, arXiv: 2206.06215

## 补充材料

**图S1** g波段定标样本在赫罗图上的分布
**图S2** SCR′方法中, 每幅图每个通道上标准星个数的分布
**图S3** SCR方法3个波段上控制样本平场改正
**图S4** 两种方法构建的流量标准星的模型星等与定标后的星等之差随消光、定标后星等和颜色的变化
**图S5** 拟合残差在CCD上的空间分布
**图S6** 流量定标一致性的内部检验
**图S7** 基于Gaia DR3无缝光谱借助XPSP方法合成的PS1星等转换后的南山星等与南山定标后星等之差在3个波段上的空间分布
**表S1** $(G_{BP}-g)_0$和$(G_{RP}-r)_0$颜色的红化系数修正项系数
**表S2** 3个波段上, SCR方法的控制样本内秉颜色拟合多项式(关于温度$T_{eff}$和丰度[Fe/H])系数
**表S3** $(G_{BP}-g)_0$和$(G_{RP}-r)_0$相对$E(G_{BP}-G_{RP})$的红化系数随温度变化的关系式系数
**表S4** 3个波段上, SCR′方法控制样本内秉颜色拟合系数

本文以上补充材料见网络版csb.scichina.com. 补充材料为作者提供的原始数据, 作者对其学术质量和内容负责.

Summary for “SAGES南山1米大视场望远镜g、r、i波段测光数据流量定标”

# Photometric calibration of the Stellar Abundance and Galactic Evolution Survey (SAGES): Nanshan One-meter Wide-field Telescope g, r, and i band imaging data

Kai Xiao[1,2], Haibo Yuan[1,2*], Bowen Huang[1,2], Shuai Xu[1,2], Jie Zheng[3], Chun Li[3], Zhou Fan[3], Wei Wang[3], Gang Zhao[3], Guojie Feng[4], Xuan Zhang[4], Jinzhong Liu[4], Ruoyi Zhang[1,2], Lin Yang[5], Yu Zhang[4], Chunhai Bai[4], Hubiao Niu[4], Esamdin Ali[4] & Lu Ma[4]

[1] *Institute for Frontiers in Astronomy and Astrophysics, Beijing Normal University, Beijing 102206, China;*
[2] *Department of Astronomy, Beijing Normal University, Beijing 100875, China;*
[3] *CAS Key Laboratory of Optical Astronomy, National Astronomical Observatories, Chinese Academy of Sciences, Beijing 100101, China;*
[4] *Xinjiang Astronomical Observatory, Chinese Academy of Sciences, Urumqi 830011, China;*
[5] *College of Artificial Intelligence, Beijing Normal University, Beijing 100875, China*
* Corresponding author, E-mail: yuanhb@bnu.edu.cn

In recent years, there have been numerous wide-field imaging surveys conducted both domestically and internationally, which have had a revolutionary impact on astronomy. Uniform and accurate photometric calibration is very challenging and plays a key role in wide-field imaging surveys. The Nanshan One-meter Wide-field Telescope, with a 1.3° field of view, was used by the Stellar Abundance and Galactic Evolution Survey (SAGES) to image 4254 square degrees of the northern sky in three broadband filters (g, r, and i). In this paper, a total of approximately 2.6 million dwarf stars were constructed as standard stars, with an accuracy of about 0.01–0.02 magnitude for each band, by combining spectroscopic data from the Large Sky Area Multi-Object Fiber Spectroscopic Telescope (LAMOST) Data Release 7, photometric data from the corrected Gaia Early Data Release 3, and photometric metallicities obtained by Xu et al. Using the spectroscopy-based stellar color regression method (SCR method) and the photometric-based SCR method (SCR′ method), we performed the relative calibration of the Nanshan One-meter Wide-field Telescope imaging data of the SAGES survey. Based on the corrected Pan-STARRS DR1 (PS1) photometry, the absolute calibration was also performed, and the transform relationship between Nanshan One-meter Wide-field Telescope calibrated magnitudes and the corrected PS1 magnitudes was given. In the photometric calibration process, we analyzed the dependence of the calibration zero points on different images (observation time), different gates of the CCD detector, and different CCD positions. We found that the stellar flat and the relative gain between different gates depend on time. Variations in the telescope’s performance can be classified as either related to sudden changes in pointing or unrelated. By analyzing the spatial distribution of the stellar flat-field, we can verify the orientation and rotation of the horizon for the Nanshan One-meter Wide-field Telescope. The amplitude of gain variation in three channels is approximately 0.5%–0.7% relative to the other channel, with a maximum value of 4%. In addition, significant spatial variations of the stellar flat fitting residual are found in all the g, r, and i filters. For each band, the significant spatial structure is similar on a daily basis but different over longer timescales (e.g., seasonally). To correct the spatial-dependence errors presented by the flat-field correction fitting residuals, we used the fitting residuals from a single day to correct all the observed images taken on that day. Using repeated sources in the adjacent images, we checked and discovered internal consistency of about 1–2 millimagnitude in all the filters. Using the PS1 magnitudes synthesized by Gaia DR3 BP/RP spectra by the synthetic photometry method, we found that the photometric calibration uniformity is about 2.4, 2.3, and 0.9 millimagnitude for the g, r, and i bands, respectively, at a spatial resolution of 1.3°. A detailed comparison between the spectroscopy-based SCR and photometric-based SCR method magnitude offsets was performed, and we achieved an internal consistency precision of about 2 millimagnitude or better with resolutions of 1.3° for all the filters. And the difference between the spectroscopy-based SCR method and the photometric-based SCR method is mainly from the position-dependent systematic error of the dust reddening map of Schlegel et al. used in SCR′ method. This work has successfully broken through the “1% accuracy bottleneck” of groundbased wide-field photometric survey photometric calibration. Our results can provide method and experience references for the photometric calibration of upcoming survey data.